# A fluctuation-sensitivity-timescale trade-off in feedback-controlled dynamics

Ka Kit Kong[1], Feng Liu[2*]

[1] *The State Key Laboratory for Artificial Microstructures and Mesoscopic Physics, School of Physics, Peking University, Beijing, 100871, China*

[2] *Key Laboratory of Hebei Province for Molecular Biophysics, Institute of Biophysics, School of Health Science & Biomedical Engineering, Hebei University of Technology, Tianjin, 300130, China*

*\*Corresponding author, email: liufeng@hebut.edu.cn*

# Abstract

Feedback control is a renowned mechanism for attenuating intrinsic fluctuation in regulatory networks. However, its impact on the response sensitivity to external signals and the response timescale, which are also critical for signal transmission, has yet to be understood. In this letter, we study a general feedback-controlled network in which the feedback is achieved by a complex interactive module. By comparing the solution of Langevin equations with and without feedback, we analytically derive a fundamental trade-off between fluctuation, sensitivity, and timescale altered by the feedback. We show that feedback control cannot infinitely suppress fluctuation without the cost of reducing sensitivity or response speed. Furthermore, the lower bound for this trade-off can be reduced up to half in non-gradient dynamical systems compared to gradient systems. We validate this trade-off as a tight bound for high-dimensional systems in nonlinear regime through numerical simulations. These results elucidate the fundamental limitation of feedback control in enhancing the information transmission capacity of regulatory networks.

Noise control is critical in signal transmission processes. While intrinsic fluctuation due to, *inter alia*, the small copy number of molecules and stochastic processes including the birth and death of molecules, is inevitable in a variety of biochemical networks [1–3], feedback control has been proposed to be a prominent mechanism in suppressing this fluctuation [4–8].

However, effective signal transmission necessitates not only small fluctuation, but also a high response sensitivity to external signals [9–14]. The relation between intrinsic fluctuation and response sensitivity limits the information transmission capacity of dynamical systems. The fluctuation-dissipation theorem (FDT) is a potential candidate for this relation. However, since the FDT is derived from near-equilibrium systems, it is still challenging to extend it to far-from-equilibrium systems [15–21]. The energy dissipation of nonequilibrium systems has been proposed to be crucial in attenuating noise while maintaining high sensitivity [22–25], thus breaking the FDT. For dynamical systems, non-gradient dynamics implies the existence of a rotational flux field [26,27], which usually breaks the detailed balance in the steady-state, leading to nonequilibrium.

Furthermore, dynamic timescale also plays a crucial role in signal transmission, especially in finite time scenario. For example, it has been proposed that increasing the dynamic timescale can exploit the temporal average effect, breaking the noise-sensitivity proportionality in considering external signal noise [28]. Moreover, the thermodynamic uncertainty relation (TUR) connects the current fluctuation to dissipation in far-from-equilibrium systems, infinite time is, however, required to achieve the predicted lower bound for fluctuation [29,30]; and a higher rate of physical processes is usually connected with larger dissipation [9,31].

In this letter, we thoroughly re-evaluate the renowned feedback control mechanism for suppressing intrinsic fluctuation. We first demonstrate that fluctuation suppression is sometimes achieved by sacrificing the response sensitivity in feedback systems, using a one-dimensional (1D) textbook toy model as an example. For general feedback control, we further identify a fundamental triplet trade-off between intrinsic fluctuation, response sensitivity, and response timescale, indicating that feedback cannot infinitely suppress fluctuation without sacrificing the other two factors. Moreover, we find that the optimal signal transmission set by this triplet trade-off varies with the degree of non-gradient and the effective dimension of the dynamics.

*A general dynamic framework of feedback-controlled networks.* We consider the signal transmission of dynamical systems described by a set of *N*-dimensional Langevin equations:

$$\frac{dx_i}{dt} = f_i(\{x_j\}, I) - \beta_i x_i + \eta_i, i = 1, \dots, N, \tag{1}$$

where $I$ is the external input signal, $f_i(\{x_j\}, I)$ is the regulatory function corresponding to the synthesis term of a regulatory network, $\beta_i$ is the decay rate, $\eta_i$ is the white noise term that arises from the intrinsic fluctuation with the amplitude: $\langle \eta_i(t)\eta_j(t')\rangle = D_{ij}\delta(t-t') = D_i\delta_{ij}\delta(t-t')$, and $D_i = (f_i + \beta_i x_i)$ [32–35], where $\langle \dots \rangle$ denotes the temporal average, $\delta(t-t')$ is the Dirac delta function and $\delta_{ij}$ is the Kronecker delta. Without losing generality, we consider the external signal to be transmitted through the first node (of which the state is $x_1$), while the other nodes compose a complex feedback module (Fig. 1), thus $\frac{\partial f_i}{\partial I} \equiv 0, \forall i \neq 1$.

The fluctuation is defined as the variance of the response at a steady-state, denoted as $\sigma^2 = \langle \delta x_1^2 \rangle = \langle (x_1 - \langle x_{1,\text{s.s.}}\rangle)^2 \rangle$ (Fig. 1), which can be derived from the Lyapunov equation [6,34–36]:

$$\boldsymbol{J\Sigma_x} + \boldsymbol{\Sigma_x J^T} + \boldsymbol{D} = \boldsymbol{0}, \tag{2}$$

where $\boldsymbol{\Sigma_x} = \langle \boldsymbol{\delta x \delta x^T}\rangle$ is the covariance matrix, $\boldsymbol{\delta x} = (\delta x_1, \delta x_2, \dots)^T$ is the deviation from the steady-state, $\boldsymbol{J}$ is the Jacobian with elements $J_{ij} = \left(\partial(f_i - \beta_i x_i)/\partial x_j\right)_{x_i = x_{i,\text{s.s.}}}$, $\boldsymbol{D}$ is the diffusion matrix with elements $D_{ij} = \langle \eta_i(t)\eta_j(t)\rangle$. The response sensitivity is defined as the steady-state responding to a small perturbation of the input signal, denoted as $\kappa = d\langle x_i\rangle/dI$ (Fig. 1), following the spirit of the FDT and the linear response theory. It can be estimated by linearizing Eq. (1) near the steady-state:

$$\kappa = -\boldsymbol{e_1^T J^{-1} e_1}\frac{\partial f_1}{\partial I}, \tag{3}$$

where $\boldsymbol{e_1} = (1, 0, \dots)^T$. The timescale is defined as the slowest relaxation timescale of the dynamics near the steady-state (Fig. 1), which is the inverse of the minimal negative real part of eigenvalues of the Jacobian, denoted as:

$$T = \frac{1}{\min(-\text{Re}(\lambda_i))}, \tag{4}$$

where $\lambda_i$ is the eigenvalues of the Jacobian $\boldsymbol{J}$. Since we assume the steady-state response is achieved by a stable fixed point, all eigenvalues of $\boldsymbol{J}$ have negative real parts and thus $T > 0$.

In general, the three factors $\sigma^2$, $\kappa$, and $T$ could have high degrees of freedom that are independent to feedback controls, e.g., the magnitude of $D_1, \frac{\partial f_1}{\partial I}$, and $\beta_1$. Therefore, to evaluate the contribution of feedback, we further normalize them by their corresponding forms under the no-feedback condition, denoted as $\sigma_0^2$, $\kappa_0$, and $T_0$ with the subscript "0" indicating the no-feedback condition. A similar protocol has been proposed recently in studying the role of feedback in adaptation [37,38]. For example, in the 1D textbook toy model ($N = 1$ in Eq. (1)), based on Eqs. (2), (3), and (4), one can find the fluctuation, sensitivity, and timescale following: $\sigma^2 = \frac{\beta_1}{\beta_1 - \partial f_1/\partial x_1} \langle x_1 \rangle$, $\kappa = \frac{1}{\beta_1 - \partial f_1/\partial x_1} \frac{\partial f_1}{\partial I}$, and $T = \frac{1}{\beta_1 - \partial f_1/\partial x_1}$, where $\frac{\partial f_1}{\partial x_1}$ denotes the strength of autoregulation. The no-feedback condition is equivalent to $\frac{\partial f_1}{\partial x_1} = 0$, thus follows:

$$\begin{cases} \sigma_0^2 &= \frac{D_1}{2\beta_1} = \langle x_1 \rangle \\ \kappa_0 &= \frac{1}{\beta_1} \frac{\partial f_1}{\partial I} \\ T_0 &= \frac{1}{\beta_1} \end{cases}. \tag{5}$$

To better isolate the impact of the feedback, the other degrees of freedom, including $D_1$, $\frac{\partial f_1}{\partial I}$, and $\beta_1$ are considered to be identical between the feedback and no-feedback conditions. Therefore, by applying the normalization scheme, the effect of feedback in 1D systems follows:

$$\frac{\sigma^2}{\sigma_0^2} = \frac{\kappa}{\kappa_0} = \frac{T}{T_0} = \frac{\beta_1}{\beta_1 - \partial f_1/\partial x_1}, \tag{6}$$

reproducing the well-known properties that negative feedback ($\frac{\partial f_1}{\partial x_1} < 0$) can suppress fluctuation ($\frac{\sigma^2}{\sigma_0^2} < 1$) and speed up the response ($\frac{T}{T_0} < 1$) [4,5]. Based on the normalization scheme, Eq. (6) also implies that fluctuation suppression ($\frac{\sigma^2}{\sigma_0^2} < 1$) always requires the same fold of sensitivity reduction in 1D systems ($\frac{\kappa}{\kappa_0} = \frac{\sigma^2}{\sigma_0^2} < 1$). This result resembles one prediction of the FDT: the steady-state responding to a conjugate signal perturbation is proportional to its intrinsic fluctuation, i.e., $\kappa \propto$ $\sigma^2$. However, as the relation of Eq. (6) cannot be generalized to high-dimensional systems, it remains unclear how general feedback affects the response sensitivity, and whether a relation between fluctuation, sensitivity, and timescale exists.

For high dimensional feedback-controlled networks following the framework in Fig. 1, their corresponding no-feedback networks are always 1D systems without autoregulation, thus the normalization factors still follow Eq. (5). To further guarantee the comparability between the no-feedback and feedback conditions in high dimensional systems, no additional source of fluctuation is considered in the feedback module, i.e., $D_i = 0, \forall i \neq 1$ in Eq. (1). Relaxing this assumption does not affect our main conclusion as a lower bound for fluctuation still remains when additional independent fluctuation is considered (see Discussion for more details).

*A triplet trade-off between fluctuation, sensitivity, and timescale in feedback-controlled networks.* Based on the introduced normalization scheme, we first focus on the potential relation between fluctuation and sensitivity for 2D systems. Under this setup, solving Eqs. (2) and (3) yields:

$$\begin{cases} \sigma^2 &= -D_1 \dfrac{J_{22}^2 + \Delta}{2\Delta\tau} \\ \kappa &= -\dfrac{\partial f_1}{\partial I} \dfrac{J_{22}}{\Delta} \end{cases}, \tag{7}$$

where $\Delta = \det(J) > 0$ is the determinant of the Jacobian, and $\tau = \mathrm{tr}(J) < 0$ is the trace of the Jacobian. Normalizing by the no-feedback condition (Eq. (5)), we find $\dfrac{\sigma^2}{\sigma_0^2} / \left|\dfrac{\kappa}{\kappa_0}\right| \geq 2 \dfrac{\sqrt{T/T_{\min}}}{1+T/T_{\min}}$, where $T_{\min}$ is the fastest timescale of the dynamics defined based on the Jacobian (see Sec. I in Supplemental Material [39] for details). This inequality, however, implies that no fundamental lower bound exists for $\dfrac{\sigma^2}{\sigma_0^2}$ given sensitivity $\left|\dfrac{\kappa}{\kappa_0}\right|$, as long as the system exhibits a timescale separation $\dfrac{T}{T_{\min}} \gg 1$ (Supplemental Material [39], Fig. S1(a)), indicating fluctuation can be suppressed infinitely without affecting the sensitivity through a feedback control in 2D systems. This conclusion is extensible to higher dimensional systems based on the validation via numerical simulations (Supplemental Material [39], Fig. S1(b)).

Though there is no fundamental constraint between fluctuation and sensitivity for 2D systems, combining the effect of feedback on the timescale (Eqs. (4) and (5)), we find a closed-form triplet trade-off between fluctuation, sensitivity, and timescale following (see Sec. I in Supplemental Material [39] for details):

$$\frac{\sigma^2}{\sigma_0^2}\frac{T}{T_0}\bigg/\left(\frac{\kappa}{\kappa_0}\right)^2 > \frac{1}{2}. \tag{8}$$

This triplet trade-off indicates that the fluctuation can only be suppressed up to half without affecting either the sensitivity or the timescale by feedback control, i.e., $\frac{\sigma^2}{\sigma_0^2} > \frac{1}{2}$ when $\frac{\kappa}{\kappa_0} = 1$ and $\frac{T}{T_0} = 1$; further suppression of fluctuation ($\frac{\sigma^2}{\sigma_0^2} < \frac{1}{2}$) always sacrifices the response sensitivity ($\frac{\kappa}{\kappa_0} < 1$) or response speed ($\frac{T}{T_0} > 1$); infinite suppression of fluctuation ($\frac{\sigma^2}{\sigma_0^2} \to 0$) without affecting the sensitivity ($\frac{\kappa}{\kappa_0} = 1$) generally requires infinitely slow dynamics ($\frac{T}{T_0} \to \infty$). Although the lower bound appears to follow $\sigma^2 \propto \kappa^2$, it cannot be explained by the error propagation since the source of fluctuation considered here is intrinsic to the system rather than inherited from the input signal. A numerical simulation of 2D systems based on the Monod-Wyman-Changeux (MWC) model [40–43] with 100,000 randomly sampled parameter sets confirms that this is a tight bound (Fig. 2(a), see Sec. IV in Supplemental Material [39] for model and parameter sampling details).

For systems with higher dimensions, numerical results show that the triplet trade-off relation (Eq.(8)) is still valid and tight (Fig. 2(b)), suggesting that Eq. (8) is a fundamental trade-off for dynamics following the feedback-controlled framework in Fig. 1.

*The optimal performance is bound by the degree of non-gradient.* While the triplet trade-off (Eq. (8)) is valid for high dimensional systems, defining a lower bound for $\frac{\sigma^2}{\sigma_0^2}\frac{T}{T_0}\big/\left(\frac{\kappa}{\kappa_0}\right)^2$ being $\frac{1}{2}$, 1D systems exhibit a worse performance following $\frac{\sigma^2}{\sigma_0^2}\frac{T}{T_0}\big/\left(\frac{\kappa}{\kappa_0}\right)^2 = 1$ (based on Eq. (6)). This may result from the fact that 1D dynamics are always gradient. To test this connection, we next focus on high-dimensional gradient systems. Since the eigen dimensions of gradient systems evolve independently under small perturbation, solving $\sigma^2$, $\kappa$, and $T$ becomes tractable for arbitrary dimensions. We find that for high dimensional gradient systems, the trade-off relation is altered as (see Sec. II in Supplemental Material [39] for details):

$$\frac{\sigma^2}{\sigma_0^2}\frac{T}{T_0}\bigg/\left(\frac{\kappa}{\kappa_0}\right)^2 \geq 1. \tag{9}$$

The lower bound is now increased to 1, indicating that the suppression of fluctuation ($\frac{\sigma^2}{\sigma_0^2} < 1$) *always* sacrifices response sensitivity ($\frac{\kappa}{\kappa_0} < 1$) or response speed ($\frac{T}{T_0} > 1$). This result implies that gradient systems perform worse than non-gradient systems regarding signal transmission. Numerical simulations of local gradient systems support that this closed-form is a tight bound in both 2D and 10D dynamics (Supplemental Material [39], Fig. S2).

So far, we have found that the lower bound for $\frac{\sigma^2}{\sigma_0^2}\frac{T}{T_0}\big/\left(\frac{\kappa}{\kappa_0}\right)^2$ in gradient and non-gradient systems is 1 and $\frac{1}{2}$, respectively. Next, we investigate how this bound relies on the degree of non-gradient of a dynamical system. However, as far as we know, no common method exists to directly quantify the degree of non-gradient using a single scalar, hence we propose one based on the symmetric-antisymmetric decomposition of the Jacobian.

The Jacobian of a dynamical system (Eq. (1)) near a steady-state can always be decomposed into a symmetric and an antisymmetric part, i.e., $\boldsymbol{J} = \boldsymbol{J}^{\mathbf{s}} + \boldsymbol{J}^{\mathbf{as}}$, where the symmetric part $\boldsymbol{J}^{\mathbf{s}} = (\boldsymbol{J} + \boldsymbol{J}^T)/2$ represents a gradient component, and the antisymmetric part $\boldsymbol{J}^{\mathbf{as}} = (\boldsymbol{J} - \boldsymbol{J}^T)/2$ represents a rotational component. For a gradient field determined by $\boldsymbol{J}^{\mathbf{s}}$, the eigen dimension with the maximal eigenvalue $\lambda_1(\boldsymbol{J}^{\mathbf{s}})$ defines its stability and is decisive: if $\lambda_1(\boldsymbol{J}^{\mathbf{s}}) < 0$, $\boldsymbol{J}^{\mathbf{s}}$ represents a stable dynamic, whose relaxation timescale is given by $\left(-\lambda_1(\boldsymbol{J}^{\mathbf{s}})\right)^{-1}$; if $\lambda_1(\boldsymbol{J}^{\mathbf{s}}) \geq 0$, $\boldsymbol{J}^{\mathbf{s}}$ represents an unstable dynamic, even though $\boldsymbol{J}$ can represent stable dynamics, and $\lambda_1(\boldsymbol{J}^{\mathbf{s}})$ corresponds to the most unstable dimension and defines its reactivity [44]. For a rotational field determined by $\boldsymbol{J}^{\mathbf{as}}$, whose eigenvalues are all pure imaginary, the eigenvalue with the maximal imaginary part $\lambda_1(\boldsymbol{J}^{\mathbf{as}})$ determines the maximal flux speed as $\text{im}(\lambda_1(\boldsymbol{J}^{\mathbf{as}})) > 0$. Accordingly, we quantify the degree of non-gradient by a non-negative scalar $\chi$ defined as:

$$\chi = \frac{\text{im}(\lambda_1(\boldsymbol{J}^{\mathbf{as}}))}{\text{im}(\lambda_1(\boldsymbol{J}^{\mathbf{as}})) - \lambda_1(\boldsymbol{J}^{\mathbf{s}})}. \tag{10}$$

A large value of $\chi$ indicates a potentially high degree of non-gradient. For gradient systems, $\text{im}(\lambda_1(\boldsymbol{J}^{\mathbf{as}})) = 0$ and thus $\chi = 0$; for systems with $-\lambda_1(\boldsymbol{J}^{\mathbf{s}}) \gg \text{im}(\lambda_1(\boldsymbol{J}^{\mathbf{as}})) > 0$, its gradient part is stable and dominant, thus the degree of non-gradient is small, i.e., $\chi \sim 0$; for systems with $\lambda_1(J^s) \gg 0$, its gradient part is highly unstable and the non-gradient part is necessary to stabilize the system, implying a high degree of non-gradient and $\chi > 1$.

Based on this quantification and combining the expression of $\sigma^2$, $\kappa$, $T$, and their corresponding normalization factors (Eqs. (4), (5), and (7)), we find that for 2D systems, the lower bound for $\frac{\sigma^2}{\sigma_0^2}\frac{T}{T_0}/\left(\frac{\kappa}{\kappa_0}\right)^2$ depends on $\chi$, following (see Sec. III in Supplemental Material [39] for details):

$$\frac{\sigma^2}{\sigma_0^2}\frac{T}{T_0}/\left(\frac{\kappa}{\kappa_0}\right)^2 \geq \mathcal{B}_G(\chi) = \frac{1}{2}\left[1 + \left(\frac{1}{1+\chi}\right)^2\right]. \tag{11}$$

For gradient systems, the lower bound is $\mathcal{B}_G(0) = 1$; for systems with a high degree of non-gradient, $\mathcal{B}_G(+\infty) \to 1/2$. Moreover, $\mathcal{B}_G(\chi) > \frac{1}{2}$ is a monotonically decreasing function, implying that a high level of non-gradient always improves the optimal performance. Numerical simulations support that Eq. (11) sets a tight lower bound for 2D systems (Fig. 3(a)) and it is also valid for high-dimensional systems (Supplemental Material [39], Fig. S3(a)).

It is worth noting that the decomposition of dynamical systems into a gradient and a non-gradient field is not unique, e.g., decomposition based on the steady-state distribution, the Helmholtz decomposition, etc. [26,45,46]. We use the symmetric-antisymmetric decomposition as it depends only on the Jacobian and thus facilitates further analytical derivation. Moreover, the rotational component of the symmetric-antisymmetric decomposition is always a curl field when projected onto any 2D subspace in which the curl operator is well-defined.

*The lower bound for the triplet trade-off is affected by the effective dimension.* While the dimension of the *de facto* trajectory of a dynamical system is usually lower than that of the whole variable space, it is critical to characterize the effective dimension of the actual dynamics because a lower bound for the triplet trade-off less than 1 necessitates a dimension higher than 1. To have an effective dimension higher than 1 requires, at least, the two slowest relaxation modes to occur at a similar timescale. This condition can be quantified by a ratio $T/T_2$, where $T$ is the slowest timescale as defined before, and $T_2$ is the second slowest timescale. In 2D systems, we find a lower bound for $\frac{\sigma^2}{\sigma_0^2}\frac{T}{T_0}/\left(\frac{\kappa}{\kappa_0}\right)^2$ depending on $T/T_2$ (see Sec. III in Supplemental Material [39] for details):

$$\frac{\sigma^2}{\sigma_0^2}\frac{T}{T_0}\bigg/\left(\frac{\kappa}{\kappa_0}\right)^2 > \mathcal{B}_\mathrm{D}\left(\frac{T}{T_2}\right) = \frac{T/T_2}{1+T/T_2}, \tag{12}$$

where $\mathcal{B}_\mathrm{D}(T/T_2) \geq 1/2$ since $T \geq T_2$. Eq. (12) implies that when $T/T_2 \to +\infty$, the lower bound $B_\mathrm{D}(+\infty) \to 1$, which is reasonable since this timescale separation implies the central manifold of the dynamic is effectively 1D. When $T/T_2 \sim 1$, the two underlying dimensions are on the same timescale, allowing the system to exploit a 2D phase space and achieve a lower bound $B_\mathrm{D}(1) = 1/2$. Numerical simulations support that Eq. (12) sets a tight lower bound for both 2D systems (Fig. 3(b)) and high dimensional systems (Supplemental Material [39], Fig. S3(b)).

Eqs. (11) and (12) appear to be two different bounds for the triplet trade-off relating to different dynamical properties of the system, they are not completely independent (Fig. 3). While a large $\chi$ ensures the potential for a high degree of non-gradient, a small $\frac{T}{T_2}$ permits the dynamical trajectory to exploit higher dimensional space and, thus, the degree of non-gradient.

*Discussion*. While noise attenuation is admittedly a crucial task in regulatory networks, the performance of signal transmission equally relies on response sensitivity, timescale, along with fluctuation, and probably more factors. In this letter, we identify a fundamental triplet trade-off in feedback-controlled dynamics (Eq. (8)), highlighting the limitation of feedback in enhancing information transmission capacity. In addition to the connection between fluctuation and response sensitivity, our results also explicitly quantify the effect of timescale, a less understood factor, resonating with the spirit of finite-time thermodynamics, in which revealing "the cost of haste" [47,48] is the central task. Moreover, we discover that the lower bound for this trade-off can be reduced from 1 to, at most, $\frac{1}{2}$ by increasing the degree of non-gradient (Eq. (11)), which is closely connected to the degree of nonequilibrium in dynamical systems. However, the trade-off (Eq. (8)) cannot be lifted by any means unlike the energy-constrained trade-offs discovered in other biological processes such as biochemical sensing [49], in which the lower bound of noise correlates with energy dissipation. Moreover, in stochastic thermodynamics, how dissipation bounds factors, such as fluctuation, response, and timescale, have been revealed over the past decades [15–17,29–31], the constraint between these factors, however, remains to be explored. Our results imply that simultaneously considering multiple factors could reveal more comprehensive constraints.

In our derivation, we focus on quantities assessing the effect of feedback, i.e., $\frac{\sigma^2}{\sigma_0^2}$, $\frac{\kappa}{\kappa_0}$, and $\frac{T}{T_0}$. This normalization scheme can exclude other effects, e.g., changing the average level $\langle x_1 \rangle$ is well known to affect the intrinsic fluctuation, changing the derivative $\frac{\partial f_1}{\partial I}$ and the decay rate $\beta_1$ can evidently alter the response sensitivity and timescale, respectively. These changes of fluctuation, sensitivity, and timescale are not unique to, even independent of, feedback control, and can be eliminated by the employed normalization scheme.

Moreover, in terms of signal transmission, this normalization can further be used to evaluate the increment of the mutual information between the input ($I$) and the output ($x_1$) (Fig. 1), which has been proposed to be effective in quantifying signal transmission capability [13,43,50–55]. Under the small noise limit and a Gaussian approximation, the mutual information between $I$ and $x_1$ follows $\mathbb{I}(I; X_1) = S(I) - S(I|X_1) = S(I) - \frac{1}{2}\left\langle \ln\left(2\pi e \frac{\sigma^2}{\kappa^2} \frac{T}{T_{\text{read}}}\right)\right\rangle_I$, where $S(I)$ is the differential entropy of the input $I$, $S(I|X_1) = \frac{1}{2}\left\langle \ln\left(2\pi e \frac{\sigma^2}{\kappa^2} \frac{T}{T_{\text{read}}}\right)\right\rangle_I$ is the conditional entropy of a Gaussian distribution, $\langle ... \rangle_I$ indicates averaging over $I$, and the factor $\frac{T}{T_{\text{read}}}$ is introduced as the noise attenuation effect due to temporal average through a reading time span of $T_{\text{read}}$, in which the correlation time of noise is approximately $T$ and $T_{\text{read}} \gg T$. Comparing with the mutual information under the no-feedback condition $\mathbb{I}_0(I; X_1) = S(I) - \frac{1}{2}\left\langle \ln\left(2\pi e \frac{\sigma_0^2}{\kappa_0^2} \frac{T_0}{T_{\text{read}}}\right)\right\rangle_I$ yields an information increment due to feedback as: $\Delta \mathbb{I} = -\frac{1}{2}\left\langle \ln\left(\frac{\sigma^2}{\sigma_0^2} \frac{T}{T_0} \big/ \frac{\kappa^2}{\kappa_0^2}\right)\right\rangle$, in which the normalization scheme naturally arises. The triplet trade-off further bounds the maximal increment of information as $\Delta \mathbb{I} \leq -\frac{1}{2} \ln \mathcal{B}$, where the factor $\mathcal{B}$ is given by the lower bound of the triplet trade-off (Eq. (8), (11), or (12)).

Several assumptions have been used in the derivation for simplification. We validate that our conclusions still hold when these assumptions are relaxed. Firstly, if the feedback module *per se* also suffers from intrinsic fluctuation, the trade-off relation (Eq. (1)) is still valid as the inclusion of any additional sources of fluctuation does not affect the lower bound. Numerical simulations support this deduction (Supplemental Material [39], Fig. S4). Secondly, our derivation is based on the small noise limit. We find that if we increase the fluctuation amplitude by 10-fold in simulations, the triplet trade-off is still valid (Supplemental Material [39], Fig S5).

Our study, as an attempt to re-evaluate the feedback mechanism in suppressing fluctuation, focuses on a specific feedback-controlled framework shown in Fig. 1. While many properties of feedback control have been revealed over the past decades in networks with relatively simple feedback pathways [5–7,12,28,35,56], biological feedback is usually more complex than being simply described as positive-feedback or negative-feedback. In our studied framework, the feedback pathway can be composed of an arbitrarily complex network; our study, thus, takes a step further into the regime of general complex feedback. One can imagine that similar relationships between the key factors controlling signal transmission could be revealed for other, even more complex, signaling frameworks in the future. These findings could help understand the real biological signaling processes, and serve as a guiding principle in designing regulatory networks for synthetic biology.

## Acknowledgments

We would like to thank Jin Wang, Yuansheng Cao, Jie Lin, Qi Ouyang, Fangting Li and Dianjie Li for their helpful discussion. This project is supported by the National Natural Science Foundation of China 32271293 and 11875076. The numerical calculation was performed on the High-Performance Computing Platform of the Center for Life Sciences, Peking University.

**Figure Captions**

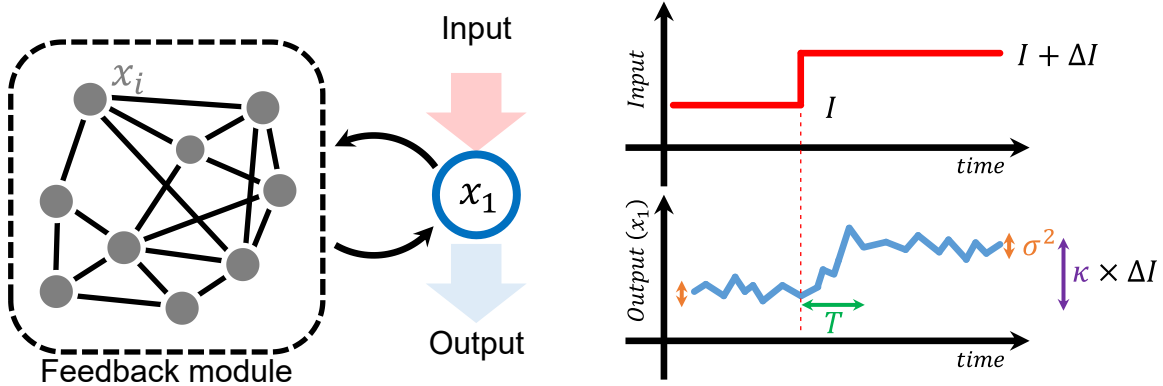

FIG. 1. The fluctuation, response sensitivity, and response timescale of feedback-controlled networks. The feedback module can be composed of multiple interactive nodes. The number and form of interactions between the feedback module and the response node ($x_1$) is not restricted. The response sensitivity ($\kappa$) is defined as the change in the steady-state average of $x_1$ under a perturbation of the input signal $I$. The response fluctuation ($\sigma^2$) is defined as the variance of $x_1$ at the steady-state. The timescale ($T$) represents the speed of relaxing to the steady-state after a perturbation, which equals the slowest mode of the dynamics.

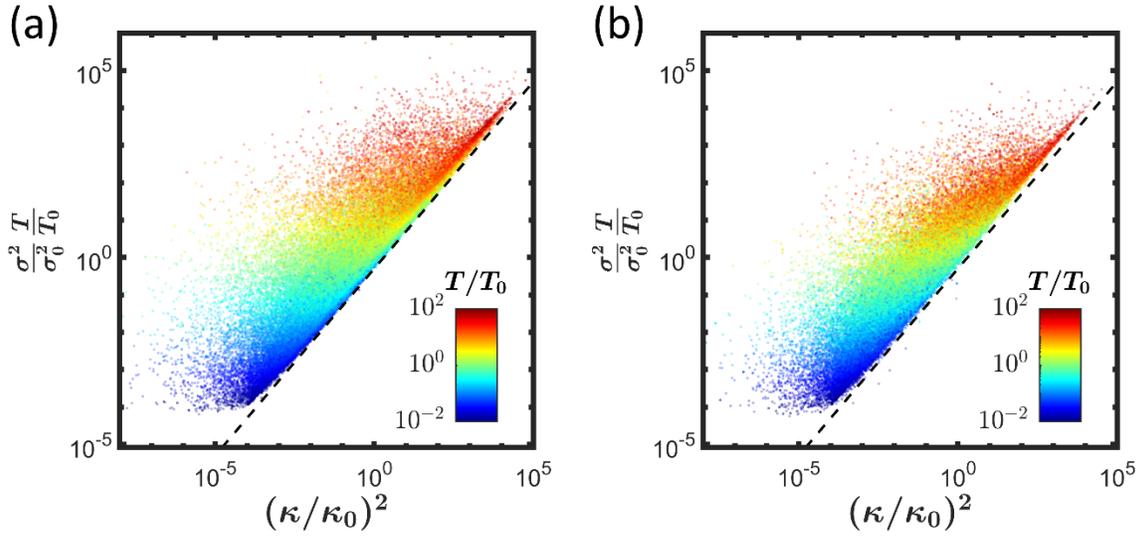

FIG. 2. The fluctuation-sensitivity-timescale trade-off in 2D and 10D dynamical systems. (a, b) Numerical results of 2D (a) and 10D (b) dynamics, in which the feedback module is composed of 1 and 9 nodes, respectively (see Sec. IV in Supplemental Material [39] for details). The results show that the predicted lower bound (Eq. (8)) is a tight lower bound (dashed line).

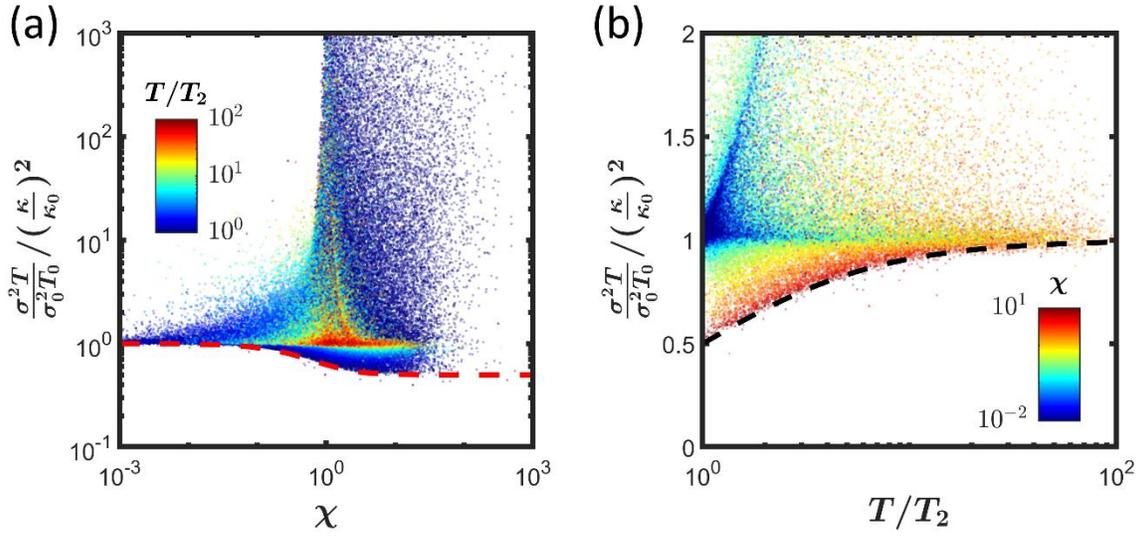

FIG. 3. The lower bound for $\frac{\sigma^2}{\sigma_0^2}\frac{T}{T_0}/\left(\frac{\kappa}{\kappa_0}\right)^2$ is determined by the degree of non-gradient $\chi$ (Eq. (10)) (a) and the ratio between the two slowest timescales $T/T_2$ (b) in 2D systems. (a, b) Numerical results of 2D systems show that the predicted lower bound in Eqs. (11) and (12) (dashed curve in (a) and (b), respectively) are tight lower bounds.